\documentstyle[preprint,aps]{revtex}
\def\ra{\rightarrow}
\tightenlines
\def\D{D\overline{D}}
\begin{document}
\preprint{\vbox{\baselineskip=14pt
\rightline{UH-511-909-98} \break
    \rightline{March 12, 1999}
  }}

\draft
\title{Implications of a $\chi$(3.86) State for Theoretical Models}

\author{S.F. Tuan}
\address{Department of Physics \\
 University of Hawaii at Manoa \\
              Honolulu, HI 96822 
\\
\vspace{.25in}
[Submitted to Phys. Lett. {\underline{B}}]}

\date{March 1999}

\maketitle
\begin{abstract}
Recent preliminary evidence for a $D \overline{D}$ state $\chi$(3.86)
raises extremely interesting implications for theoretical models,
particularly those currently in vogue, e.g. the flux tube model and a
molecular charmonium picture.  We examine the experimental consequences
as well as the possible unveiling of a strata of C-exotic states in
heavy quarkonia systems by future facilities like BES Upgrade/Tau-Charm
Factory and CLEO III/B-Factories.

 \end{abstract}

\pacs{ }

In a thought provoking Ph.D thesis Ms. Lingjun Pan\cite{1} studied D
meson production (at BES) in $e^+e^-$ annihilation at $E_{cm}$ = 4.03
GeV.  The shape of the D meson momentum spectrum in the 0.3 $<p_D <$ 0.7
GeV region indicates the presence of an important $D\overline{D} \pi$
production mechanism in addition to $e^+e^- \longrightarrow D^*
\overline{D}$.  She proposed that the data are best explained by the
decay chain $\psi (4.03) \longrightarrow \chi(3.86) \pi$, where
$\chi(3.86)$ denotes a narrow p-wave $\D$ molecular state with mass
near 3.86 GeV that decays via $\chi(3.86) \ra \D$.  Such a state was
predicted in 1977 by De R\'{u}jula, Georgi, and Glashow \cite{2} in
a paper that interpreted the $\psi$(4.03) (now called $\psi$(4.04) by
PDG \cite{3}) resonance as being a four-quark $D^{*} \overline{D}^{*}$
molecular state.  In this note we shall follow the convention of PDG
\cite{3} rather than \cite{1}, namely $\psi (4.04), \psi (4.160)$, and
$\psi (4.415)$ with $J^{PC} = 1^{--}$ have unassigned $I^G = ?^?$, be
extended to this new $\chi$(3.86) state also.  For instance the popular
perception that $\psi (4.04)$ is I=0  $c \bar{c}$ state would lead to the
produced $\chi$(3.86) $\D$ state having isospin 1 thus precluding it
from being a $c \bar{c}$ charmonium state.

In view of the central role of $\psi$ (4.04) to interpretation, let us
review first the current experimental facts known about this state. The
PDG \cite{3} states that $\psi$ (4.04) is $I^G = ?^?$, $J^{PC} = 1^{--}$
with mass $4040 \pm 10$ MeV, full width $\Gamma = 52 \pm 10$ MeV, and
$\Gamma_{ee} = 0.75 \pm 0.15$ KeV.  These are the hard facts.  The
qualitative soft facts are that $\psi (4.040) \ra D^0 \overline{D}^0$,
D* (2007) $\overline{D}^0$ + c.c., and $D^* (2007) \overline{D^*} (2007)$
are all described as ``seen''.  However PDG \cite{3} allowed that a very
early 1977 Mark I Collaboration experiment by Goldhaber {\em et al.} \cite{4}
arrived at the following conclusion (with p-wave phase space factor
removed):-
\begin{eqnarray}
\Gamma (D^0 \overline{D}^0)/\Gamma (D^{*} (2007)^{0} \overline{D}^{0} +
c.c) & = 0.05 \pm 0.03 
\nonumber  \\
\\
\Gamma (D^*(2007)^0 \overline{D}^*(2007)^{0}/\Gamma (D^{*} (2007)^{0} \overline{D}^{0} +
c.c) & = 32 \pm 12.0. \nonumber 
\end{eqnarray}

Equation (1) is consistent with molecular charmonium expectations
\cite{2} and in fact motivated the theoretical speculation
(Ref. \cite{2} is qualitative to the extent that no model calculations
were done).  Though De R\'{u}jula {\em et al.} \cite{2} were aware of
the $\psi$ (4.415) peak, now known \cite{3} to have $I^G = ?^?, J^{PC} =
1^{--}$ with {\em narrow} width $\Gamma = 43 \pm 15$ MeV, dominant decay into
hadrons, and $\Gamma_{ee} =0.47\pm 0.10$ KeV, they were unaware of the $\psi
(4.160)$ with $I^G = ?^?, J^{PC}= 1^{--}, \Gamma = 78 \pm 20$ MeV and
$\Gamma_{ee} = 0.77 \pm 0.23$ KeV.  Given the near `degeneracy' between
$\psi$ (4.040) and $\psi(4.160)$, perhaps they can be regarded as the
$I^G = 0^-, 1^+$ members of a p-wave $D^* \overline{D}^*$ short-lived
four-quark ``molecule'' in which the $D^*$ and $\overline{D}^*$ maintain
their ``atomic'' integrity. The $\psi$ (4.415) is then understood as an
$S$-wave bound state of a $D$ or $D^*$ with a p-wave $\overline{D}^{**}$
as originally proposed \cite{2}.  This interpretation allows the newly
found $\chi(3.86)$ \cite{1} to be the narrow $\D$ p-wave molecular
charmonium state $I^G = 1^+, J^{PC}  = 1^{--}$ if $\psi (4.04)/ \psi
(4.160)$ is $I^G = 0^-$ and $I^G = 0^-$, $J^{PC}= 1^{--}$ if $\psi
(4.04)/\psi (4.160)$ is $I^G = 1^+$.

However it has been pointed out by Close and Page \cite{5} that the
lowest hybrid $c{\bar c}g$ charmonia ``$H_c$'' {\em predicted} by
flux-tube charmonium spectroscopy are in the 4.1 - 4.2 GeV region, while
the QCD inspired potential models with long range linear behavior
uniformly predict that the $\psi$ (3S) $c{\bar c}$ state is also in the
narrow range 4.10 to 4.12 GeV. Philip Page has shown \cite{6}
that within the conventional charmonium picture, the hadronic width and
branching ratios of $\psi$ (4.040) and the narrowness of the $\psi$
(4.415) constrain the $\psi$ (4.040) to be the 3S $c{\bar c}$ state and
the $\psi$ (4.415) to be very likely 4S $c{\bar c}$ state. Hence Close
and Page \cite{5} propose an attractive {\em quantitative} picture, in
which the $\psi$ (3S) $c{\bar c}$ and hybrid $H_c$ with content $c{\bar
c}g$ are within 30 MeV of each other and, again, in the 4.1 to 4.2 GeV
region. Within their widths, there will be mass degeneracy of $\psi$
(3S) and $H_c$, leading to strong mixing and splitting of the
eigenvalues. If such a degeneracy occurs one immediately expects that
the physical eigenstates will tend to be
\begin{equation}
\psi_{\pm} \simeq \frac{1}{\sqrt{2}} (\psi(3S) \pm H_c)
\end{equation}
where $\psi_{-} \equiv \psi$ (4.040) and $\psi_{+} \equiv \psi$
(4.160). The $H_c$ component of Eq.(2) will be ``inert'' because a
characteristic feature of hybrid mesons in the flux-tube model is the
prediction that their decays to ground state mesons (e.g. $c\bar u$,
$c\bar d$, $c\bar s$ + c.c., L=0) are suppressed and that the dominant
coupling is to excited states (as orginally suggested by Isgur {\em et
al.} \cite{7} in the light quark context), in particular to $D\overline{D}^{**}$ for
which the threshold is $\sim$ 4.3 GeV. This pathway will hence be closed
for $\psi$ (4.040) and $\psi$ (4.160) and consequently the hadronic
decays of $\psi_{\pm}$ will be driven by the $\psi$ (3S)
component. Close and Page \cite{5} were able to obtain a {\em
quantitative} understanding of the known experimental features (total
widths, leptonic branching ratios, and hadronic decays) of {\em both}
$\psi(4.040) \equiv \psi_{-}$ and $\psi(4.160) \equiv \psi_{+}$ on this
basis.

Clearly the two interpretations for the $\psi$ (4.040), $\psi$ (4.160),
and $\psi$ (4.415) states are in many ways orthogonal to one
another. The flux tube interpretation \cite{5} seems to be on a more
quantitative basis, while the molecular charmonium interpretation,
though speculative, has intuitive appeal. For instance in the light
quark system it is believed \cite{8} that the $f_o$ (980) and $a_o$
(980) with $J^{PC}$ =  $0^{++}$ and $I^G$ = $0^+$ and $1^-$ respectively
are essentially bound s-wave $K\overline{K}$ molecules. The D's should
interact among themselves much as the K's do. Because the D's are much
heavier than the K's it is even more likely that there should exist
molecules made up of two oppositely charmed mesons. {\em  A key
experimental test to differentiate between molecular charmonium and flux
tube model [$c\bar c$g + $c\bar c$/$\psi$ (3S)] is that the $I^G$ =
$1^+$, $J^{PC}$ = $1^{--}$ $\psi$ (4.040), $\psi$ (4.160), $\psi$
(4.415) would not exist for the flux tube case.} Hence experimental
effort should be devoted towards identifying the decay chains given in
Fig. 2(b) and Fig. 3(b) of Ref. [2].

In terms of the proposed $\D$ state $\chi$(3.86) \cite{1} that is
strongly produced via $\psi (4.040) \longrightarrow \pi + \chi(3.86)$,
the flux-tube approach \cite{5} where $\psi$ (4.040)/$\psi$ (4.160) must
be $I=0$, {\em forces $\psi$ (3.86) to be $I=1$.} On the otherhand in
molecular charmonium picture \cite{2}, the state $\chi$(3.86) seen
could correspond to either $J^PI^G = 1^-1^+$ P-wave $\D$ molecular state
from $\psi$ (4.040)/$\psi$ (4.160) with $I^G$ = $0^-$ or its $\D$
$1^-0^-$ counterpart from $\psi$ (4.040)/$\psi$ (4.160) with
$I^G=1^+$. Establishment of $\chi$(3.86) as $I=0$ would imply that one
of $\psi$ (4.040)/$\psi$ (4.160) is $I=1$, thus ruling out the flux-tube
interpretation. The decay width of such a p-wave state just above the
$\D$ mass threshold would be suppressed by the limited phase space and
an angular momentum barrier, making it reasonable to expect that it
would be rather narrow.

To the extent that spin/parity has not been carried out on $\chi$
(3.86), we must consider other spin/parity assignments. Molecular
charmonium model \cite{2} also predicts s-wave $J^P=1^+$ and $J^P=0^+$
molecular states. The $J^P=1^+$ is ruled out for $\chi$(3.86) since 
$1^+ \not \leftrightarrow \D$. Such an s-wave $J^P = 1^+$ 
state if below the $D^*\overline{D}$ mass threshold 
(3.871 to 3.879 GeV) would decay via cascades through the $\psi$ (2S)
and $J/\psi$ 
production at the $\psi$ (4.03) \cite{9}, which indicates 
that such states, if they exist, are above 
$D^*\overline{D}$ mass threshold. Above $D^*\overline{D}$ threshold, 
D-mesons from their decay would populate the $0.0 < p_D < 0.3$ GeV/c 
part of the momentum region, where there are large numbers of 
D mesons from $D^*\overline D^*$ production \cite{1}. 
Hence it is not clear that such a $J^P=1^+$ molecular state 
would have been found in the experiment at hand \cite{1}. 
Molecular s-wave states with $J^P=0^+$ are expected with masses 
below those of the corresponding p-wave states, indeed when two 
particles almost bind in a p-wave state, it is virtually certain 
that they do form an s-wave bound state \cite{2}. Hence even more 
so than the $K\overline{K}$ molecules $f_o(980)/a_o(980)$ \cite{8}, 
we expect the $J^P=0^+$ molecules of molecular charmonium 
to be {\em below} the $\D$ threshold, hence 
not relevant for $\chi$(3.86) interpretation. On the otherhand 
an s-wave $J^P=0^+$ state below the $\D$ mass threshold would 
decay via cascade transitions to the $\eta_c$. Such decays would 
be hard to observe in BES, and thus could easily escape detection. 
In a purely $c\bar c$ charmonium framework $\chi_c^{\prime}$ states 
are predicted \cite{10} with masses $2^{3}P_{0}$ (3.92), 
$2^{3}P_{1}$ (3.95), $2^{3}P_{2}$ (3.98). The $2^{3}P_{0,2}$ 
$\chi_c^{\prime}$ states lie considerably higher in mass than 
$\chi$(3.86); furthermore a $c\bar c$ interpretation for 
$\psi(4.03)/\chi(3.86)$ would forbid $\psi(4.03) \rightarrow 
\pi + \chi(3.86)$. Given the uncertainty in mass predictions, 
it seems premature to assess whether current BES capabilities
could/could not have explored for these $\chi_c^{\prime}$ states.

The recent study \cite{1}, in arriving at the $\D$ resonant
interpretation 
for the anomaly at 3.86 GeV, had ruled out the effect being due to (i)
instrumentation, (ii) wrong $D^{*+}$ branching fractions, (iii) isospin
violations in $D^*\overline{D}$ production, (iv) non resonant
$D\overline{D}\pi$ production, and (v) additional $D\pi$ resonances
\cite{11}. Given the importance of this state for theory, more data are
needed to verify the existence of such a state. A ten times larger event
sample would provide enough double-tag events to start understanding the
properties of this state. In this case it may be possible to isolate
a sample of enough events where both D mesons have a momentum that is
distinct from the expected bachelor D momentum so that a peak in the
$\D$ invariant mass plot could be apparent. Of particular interest is
the charge structure of the anomalous events, i.e. the relative numbers
of $D^o\overline{D^o}$, 
$D^+\overline{D^o}$, and $D^+D^-$ tags, would tell us 
{\em the isospin of the state} critical to theoretical interpretations \cite{2,5}.  

Irrespective of which of the two contending models \cite{2,5} survive 
the test of future experiments, it is now completely clear that {\em
both} models support a strata of C-exotic states for heavy quarkonia
systems at masses comparable to what has been discussed above (together
with their counterparts in the $B\overline{B}$ system). In the flux-tube
version \cite{5}, if the C-exotic $1^{-+}$ state lies below the $\psi$
(4.040) or $\psi$ (4.160), there is the possibility of a direct spin flip
$M1$ radiative transition of the 
hybrid component $H_c \rightarrow \gamma 1^{-+}$ exposing the 
C-exotic hybrid unambiguously. The matrix element for $H_c \rightarrow 
\gamma 1^{-+}$ is related to that of $J/\psi \rightarrow \eta_c \gamma$ 
in the limit where the photon momentum tends to zero. As shown 
explicitly \cite{5}, the relative rates scale as (where p stands for the
momentum of the photon in the rest frame of $\psi_\pm)$ 
\begin{equation}
\Gamma(\psi_{\pm} \rightarrow \gamma H_c(1^{-+})) \sim 0.3
\left ( \frac{p}{118 \ MeV} 
\right )^3 KeV
\end{equation}
implying a branching ratio
\begin{equation}
B(\psi_{\pm} \rightarrow \gamma H_c(1^{-+})) \sim 10^{-5}.
\end{equation}
The subsequent transition
\begin{equation}
\psi_{\pm} \ra H_c(1^{-+}) + \gamma \rightarrow J/\psi(3.095)\gamma\gamma
\end{equation}
may provide the $J/\psi(3.095)$ as a tag, though a dedicated search at a
high intensity Tau Charm Factory, given the type of branching ratio (4),
may still be required to isolate this signal. There may be analogous
signals in the $\Upsilon$ spectrum. There could be a mass shift of
$\Upsilon$ (4S) [usually identified with $\Upsilon$ (10.580)] and the
candidate $\Upsilon$ (5S)(10.580) that appears qualitatively similar to
those of $\psi$ (4.040) and $\psi$ (4.160) which can conceivably be
explained by an increase of 80 MeV in the $\Upsilon$ (5S) - $\Upsilon$
(4S) mass separation due to coupled channel effects. Given that there
will soon be extensive studies in the 4S peak at CLEO III/B-Factories it
may be interesting to see if there are any anomalous radiative decays 
to $1^{-+} \ H_b$ state arising from a possible $H_b$ component mixed into 
the 4S peak. Such a $H_b$ hybrid $b\bar{b}g$ state with
$I=0,J^{PC}=1^{-+}$ 
has been predicted by P. Hasenfratz {\em et al.} \cite{12}, 
on the basis of the bag model, with mass value 10.49 GeV.  More recent
study
[K.J. Juge et al., hep-ph/9902336] using numerical lattice type
simulation, gives lowest $b\overline{b}g$ hybrid mass of about 10.9 GeV - above $B
\overline{B}$ threshold near $\Upsilon (4S)$.  However lattice work have
a tendency towards giving higher mass values.

As concluded \cite{2}, the molecular charmonium picture provides us 
with a spigot to a fascinating but otherwise almost inaccessible new 
``molecular'' spectroscopy full of experimental and theoretical
challenges. 
Never is the spigot more compelling than in the arena of unveiling a new 
strata of C-exotic mesonic states. Long before the advent of hybrid
systems with gluon content (which allows for C-exotic states), 
it was known that C-exotic mesons (called C-abnormal mesons by Gell-Mann 
\cite{13}) might exist, not coupled to $N\overline{N}$, $q\bar{q}$ and 
hence have {\em low production rates} in many processes. This can be 
attributed, at least partially, to the absence or smallness of many pole 
terms (one-particle exchange) that are fully allowed in the case of 
normal mesons. The work of De R\'{u}jula {\em et al.} \cite{2} is 
especially interesting in that they suggest that {\em P-wave four-quark 
system can lie close to threshold in the charm system reached by
$e^+-e^-$ annihilation}, because of the heaviness
 of the D's (ipso forte for the B's). It has long been 
recognized that though C-exotic mesons cannot be formed from the 
$q\bar{q}$ system with L excitations, nor with the s-wave states 
formed out of $q\bar{q}q\bar{q}$ configuration \cite{14}, 
{\em the low-spin ones can be formed as P-wave states out of 
a four-quark configuration}. Hence {\em they should lie at
 mass comparable to that of the $\psi$ (4.040) state} \cite{15}. 
The prominent C-exotic states are
\begin{eqnarray}
J^{PC} & = 0^{--}; I^G = 0^-, 1^+\nonumber\\
J^{PC} & = 0^{+-}; I^G = 0^-, 1^+\nonumber\\
                                          \\
J^{PC} & = 1^{-+}; I^G = 0^+, 1^-\nonumber\\
J^{PC} & = 2^{+-}; I^G = 0^-, 1^+.\nonumber
\end{eqnarray}
With our concentration on $\D$, $D\overline{D^*}$, 
$D^*\overline{D^*}$ systems for molecular charmonium, $I=0$ and 
$I=1$ channels are in any case the appropriate ones to emphasize.

The strong-decay pattern of C-exotic $\psi$ particles of Eq. (6) into
($\D$, $D\overline{D^*}$, $D^*\overline{D^*}$) can be summarized 
as follows. All combinations of $J^{PC}$ in (6) do not decay into 
$\D$ because of C conservation for $J^{PC} = 0^{--}, 1^{-+}, 2^{+-},$ 
and P conservation for $J^{PC} = 0^{+-}$. The$J^{PC} = 0^{--}, 1^{-+}$ 
members have strong decays via the P-wave into ($D\overline{D^*} \pm 
\overline{D}D^*$), while those of $J^{PC} = 0^{+-}, 2^{+-}$ have strong
decays 
via the D-wave into ($D\overline{D^*} \pm \overline{D}D^*$). Finally, 
the $D^*\overline{D^*}$ mode is forbidden for $J^{PC} = 0^{--}, 0^{+-}$
 because of C invariance while allowed for $J^{PC} = 1^{-+}, 2^{+-}$ 
as P- and D-wave molecules, respectively. Clearly, the low-lying 
C-exotic molecular charmonium states are likely to be the P-wave molecules
\begin{eqnarray}
\psi(J^{PC} & = 0^{--}; I^G = 0^-, 1^+),\nonumber\\
                                                 \\
\psi(J^{PC} & = 1^{-+}; I^G = 0^+, 1^-),\nonumber
\end{eqnarray}
which may be located at a mass comparable to the P-wave ($J^{PC}=1^{--}; 
I^G=0^-,1^+$) $D^*\overline{D^*}$ resonance interpretation for the $\psi 
(4.040)/\psi (4.160)$ structure in $e^+-e^-$ annihilation. Since it
seems 
more favorable to exploit the discovery of the C-exotics as an end
product 
of cascade chains from a {\em higher-mass $J^{PC}=1^{--}$ molecule}, 
in Fig. 1 of paper \cite{15} the appropriate transitions from 
isoscalar/isovector $\psi$ (4.415) to the C-exotics of Eq. (7) 
are delineated in some detail.

We need to note though that for normal mesons made from light u,d,s
quarks, 
such a P-wave four-quark system may lie sufficiently higher than the 
s-wave $K\overline{K}$ molecules as to be less tractable
experimentally. 
Hence C-exotics from P-wave four (light)-quark systems are likely to be 
significantly further away from the appropriate thresholds, as appears
 also to be the case for hybrid C-exotics \cite{16}.

This work was supported in part by the U.S. 
Department of Energy under Grant DE-FG-03-94ER40833 at the 
University of Hawaii-Manoa.  Helpful discussions with P.R. Page, Lingjun
Pan, and S. L. Olsen are also acknowledged.

\end{document}